# SR-DEM: an efficient discrete element method framework for particles with surface of revolution

Fei-Liang Yuan (FeiLiang.Yuan@gmail.com)[1]

[1] Multiphase Flow Systems, Institute of Process Engineering, Otto-von-Guericke-University Magdeburg, Hoher Weg 7b, D-06120, Halle (Saale), Germany

**Abstract**

In this paper, we present the surface of revolution discrete element method (SR-DEM) to simulate systems of particles with closed surfaces of revolution. Due to the cylindrical symmetry of a surface of revolution, the geometry of any cross-section about the axis of rotation remains the same. Taking advantage of this geometric feature, a node-to-cross-section contact algorithm is proposed for the efficient contact detection between particles with surface of revolution. In our SR-DEM framework, the contact algorithm is realized in a master-slave fashion: the master particle is approximated by its surface nodes, while the slave particle is represented by a signed distance field (SDF) of the cross-section about the axis of rotation. This hybrid formulation in both 2D and 3D space allows a very efficient contact calculation yet relatively simple code implementation. We then apply SR-DEM to simulate particle-particle, particle-wall impact, granular packing in a cylindrical container and tablets in a rotating drum, to demonstrate SR-DEM's ability to predict the post-impact velocities, packing porosity and dynamic angle of repose, respectively. Finally, we suggest a simple approach to find an optimal surface resolution, by increasing the number of surface nodes until some of the bulk properties that could characterize the system converge.

# 1. Introduction

Granular materials encountered in many industrial processes are mostly non-spherical. Since the shape of non-spherical particles plays an important role on the dynamics of granular flows [22, 25, 47, 54], ignoring the shape effect and using spherical particles in simulations usually does not yield satisfactory results. Data obtained from scaled laboratory experiments may provide limited information that characterizes the granular systems, because it is difficult to measure the microscopic flow behavior, due to the opaque nature of solid particles. In industry, the Discrete Element Method (DEM), firstly introduced by Cundall and Strack [7], is commonly applied to get a better understanding of how to optimize granular flow processes, e.g. mixing, coating, conveying, sieving, crushing and milling, as numerical results from DEM simulations provide both micro- and macroscopic "measurements", where the particles with realistic shapes can be properly modeled.

For simplicity, a particle with closed surface of revolution (SR) is termed as a SR particle, its shape is termed as SR shape, and the cross-section of a SR particle about the axis of rotation is termed as major cross-section in this paper. Previously, several dedicated contact algorithms have been developed for specific particle shapes that can be cataloged in the family of SR particles. Examples include ellipsoidal particles [2, 26], tablet-shaped particles [30, 55], cylindrical particles [15, 19, 29] and sphero-cylinders [33, 59]. These approaches are usually accurate and efficient in the contact detection, as the particle shapes are modeled by real (implicit) geometries. Apparent drawbacks of these approaches are: (1) particles are limited to one kind of shape, and (2) the numerical implementation can be complicated.

Some super-ellipsoids or super-quadrics [4, 16, 40, 49], whose shapes are described by the formula $(|\frac{x}{a}|^{2/e} + |\frac{y}{b}|^{2/e})^{e/n} + (|\frac{z}{c}|)^{2/n} = 1$, can also represent ellipsoidal, cylinder-like particles. A convex surface of revolution is guaranteed by setting the blockiness parameters $e = 1$, $n \in (0, 2]$, and the half-length of the two principal axes along X- and Y-axis be equal (i.e. $a = b$). Similar to a bi-convex tablet that can be formed by three parts: two cap surfaces (spherical portion) and one cylindrical band [30], some tablets and capsules with surface of revolution can be formed by the combination of three intersecting super-quadrics [39], e.g. two oblate spheroids and one cylindrical band for bi-convex tablets; two spheres and one cylindrical band for capsules. Nevertheless, only ellipsoidal, cylinder-like and box-like particles can be modeled by super-quadric DEM. Besides, the contact calculation between super-quadric particles requires non-linear and iterative schemes (e.g. Newton-Raphson method), which are usually computationally expensive, and the convergence properties decrease as the particles become less like ellipsoids with increasing blockiness parameters ($e$ and $n$), and the aspect ratios deviating from one [5, 41]. Other similar approaches like poly-ellipsoids [48] and poly-superquadrics [57, 64] that are composed of eight pieces (octants) of ellipsoids or super-quadrics, can also produce some SR particles with proper parameters in the surface functions, but still bear similar drawbacks encountered in the super-quadric DEM.

Modeling arbitrary SR particles either convex or non-convex will be a challenging task with aforementioned methods, as they are limited to certain shapes. Versatile methods capable of arbitrary shape description may be the choice. One straightforward approach is the polyhedral DEM, where the particle surface is directly approximated by its triangle mesh. The contact detection between polyhedral particles are usually complex and expensive, given the fact that many algorithms have been reported in the literature [e.g. 3, 6, 9, 13, 24, 34, 43, 44, 50, 61], because we have to deal with the contact interactions between the surface elements in a contact pair: vertices, edges and faces (i.e. triangles) with total six different combinations of contact types. Polyhedral DEMs are efficient to model blocky particles with sharp edges, but have difficulty with smooth particles, as a large number of triangles are needed to mimic the smooth surface.

Another approach to approximate realistic particles is the mutli-sphere (MS) DEM [e.g. 12,

23, 38], which is still very popular in the DEM community, due to its efficiency in the contact detection and simplicity in practical implementation. MS particles are usually constructed by "clumping" multiple overlapping spheres together. However, how to find as few spheres as possible for a good approximation of complex shapes is a highly non-trival task [14, 17, 36, 60]. The component (primary) spheres in a MS particle are fixed in the body frame, i.e. their relative positions do not change during contacts. Multi-sphere approach provides great versatility in particle shape representation especially for smooth particles, whereas many spheres are required for particles with sharp edges. Because the shape of a MS particle is literally the boolean union of all primary spheres, the resulting surface is usually rather bumpy, thus a pair of convex particles may have multiple contact points during collision. Therefore care must be taken in the contact force calculation [1, 21, 28, 31]. For detailed reviews on the contact algorithms for non-spherical particles one might refer to the literature [41, 42, 66].

The level set based DEM (LS-DEM) recently developed by Kawamoto et al. [27] also offers good versatility in real shape description, and relatively simple contact detection with a node-to-surface algorithm commonly used in the finite element method [35]. In LS-DEM, particle surface is reconstructed by level set functions via interpolation [e.g. 37, 62], hence the implicit 3D surface can be expressed by zero level set: $S = \{\phi(\boldsymbol{x}) = 0 \mid \boldsymbol{x} \in \mathbb{R}^3\}$, where the signed distance $\phi(\boldsymbol{x})$ is zero for any point $\boldsymbol{x}$ on the particle surface. In practical implementation, the minimum axis-aligned bounding box (AABB) that tightly encompass a particle, is discretized into uniform grids. The signed distance for each grid point is simply obtained by computing the shortest distance from the grid point to the particle surface. Signed distance is negative if the grid point is inside the particle and positive outside. The node-to-surface algorithm is realized by a master-slave fashion: the master particle is represented by its surface nodes, while the slave particle is the reconstructed signed distance functions stored on the AABB grid points. If the master and slave particles are in contact, one or more master nodes may be found inside the slave particle surface. Overlap distance evaluation from master node to slave surface works essentially as lookup tables: first find the grid cell (index) that contains the master node, then the signed distance of the master node can be obtained by trilinear interpolation on the cell's grid points [27]. We can see that the contact detection in LS-DEM is quite simple, but at the price of great computational cost and memory (RAM) usage. An uniform grid with an intermediate resolution 50×50×50 yields a lookup table of size $1.25 \times 10^5$, it will be quite expensive to locate the grid cell index of a master node. The precision of LS-DEM is highly dependent on the grid resolution and number of master nodes. A good precision usually requires a few thousand of nodes, and the ratio of particle's characteristic diameter to grid cell size 20~50. Even though LS-DEM is slower than classic DEM, the OpenMP scalability in LS-DEM seems greater than in the latter [10].

The objective of this study is to develop and validate a surfaces of revolution DEM (SR-DEM) framework, which shares some similar features in the LS-DEM. The key differences and novelty here are: (1) The implicit surface is reconstructed by signed distance functions from a SR particle's major cross-section, thanks to the cylindrical symmetry. Only a few hundreds of grid points are required for a very accurate shape approximation, whereas at least tens of thousand grid points are required for a coarse or intermediate surface resolution in the LS-DEM. Furthermore, if a SR particle's major cross-section only consists of few curves that can be described mathematically, it is usually more efficient to use the explicit geometry instead in the narrow phase contact detection, thus the reconstructed cross-section via a signed distance field is not necessary. Nevertheless, node-to-curve algorithms must be developed. (2) A node-to-cross-section contact algorithm is employed for the contact detection, instead of the expensive node-to-surface approach; (3) Directly computing the distance from a node to the slave's cross-section is also efficient as an accurate (half) cross-section can be formed by 100~150 line segments. Because of the shape and contact handling in 2D space, SR-DEM offers accurate shape representation with minimum elements (grid points or line segments) yet efficient DEM framework for granular systems.

## 2. Surface of revolution DEM

### 2.1 Particle shape and properties

A surface of revolution is created by rotating a plane curve around an axis of rotation that lies on the same plane as shown in Figure 1. Suppose the red curve BC can be described mathematically by an equation $x = f(z)$ in the XZ-plane, rotating it around the Z-axis 1 revolution will yield a surface of revolution. If the curve end points do not lie on the Z-axis (rotation axis), we can project the curve end points B, C to the Z-axis, in order to obtain the projected points A, D respectively. With these two additional line segments AB, CD and the curve BC, a closed surface of revolution is generated as shown in Figure 1.

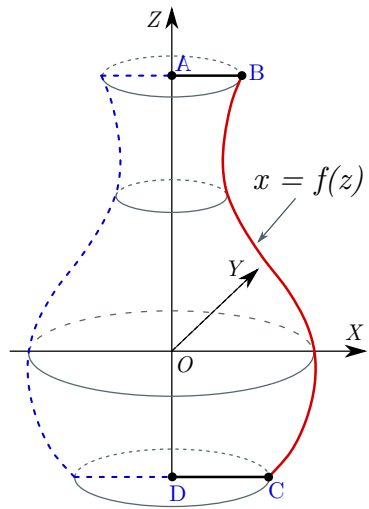

**Figure 1:** Closed surface of revolution by rotating continuous curve ABCD about the Z-axis. The curve end points must be lie on the axis of rotation, and not intersect by itself.

Since the plane curve BC can be arbitrarily shaped, a wide range of particle shapes can be covered. Sphere, spheroid, cylinder and conical frustum, just name a few. Show cases of common SR particles are plotted in Figure 2. Properties of a homogeneous SR particle such as momentum of inertia, mass and center of mass can be calculated from the curve function $x = f(z)$. As a SR particle can be seen as a stack of many thin disks with radius $R = x = f(z)$ through the Z-axis, above mentioned properties are integrated over these thin disks.

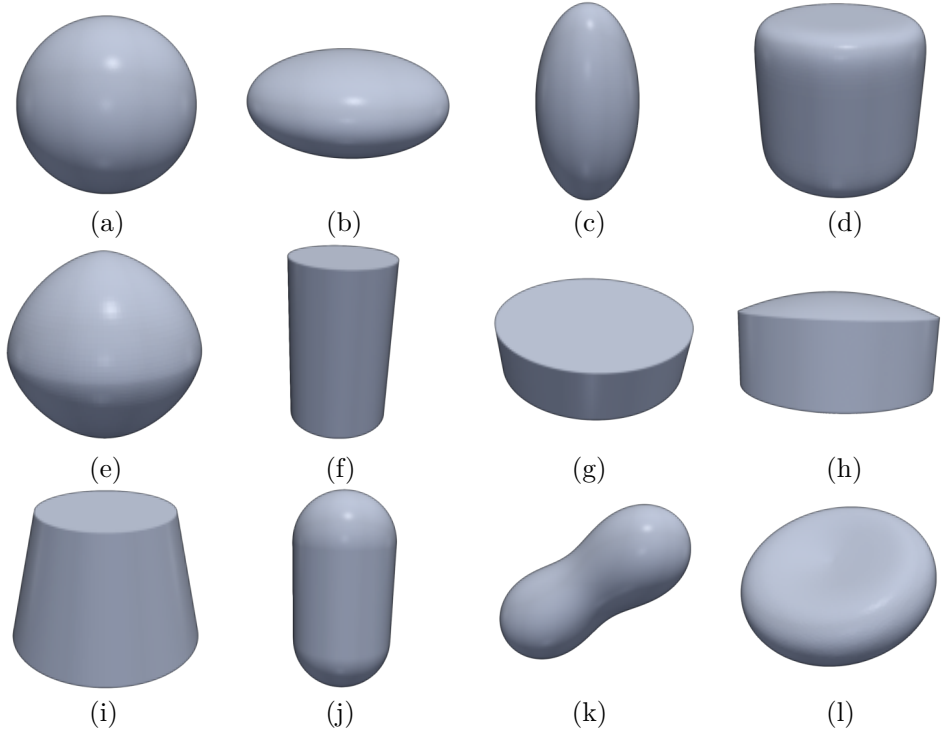

**Figure 2:** Examples of common SR particles. (a-e) shapes that can be described by super-quadric equations; (f-g) cylinder and disk; (h) bi-convex tablet; (i) conical frustum; (j) sphero-cylinder; (k) peanut shaped particle from Cassini oval; (l) red blood cell.

Particle mass with uniform density $\rho$:

$$M = \rho V = \rho \pi \int_{z_D}^{z_A} f^2(z) dz \tag{1}$$

Here $z_A$ and $z_D$ are the z-coordinates of the curve end points A and D. Due to the cylindrical symmetry of SR particles about the Z-axis, the center of mass should lie on the Z-axis, i.e. $C = (0, 0, C_z)$. $C_z$ can be calculated by:

$$C_z = \int_{z_D}^{z_A} z f^2(z) dz \bigg/ \int_{z_D}^{z_A} f^2(z) dz \tag{2}$$

Likewise, the off-diagonal coefficients in the particle's inertia tensor should be zero. Therefore, the particle's momentum of inertia can be expressed by $I = [I_x, I_y, I_z]$. The principal moments of inertia are calculated as follows [8].

$$I_z = \frac{\pi}{2} \rho \int_{z_D}^{z_A} f^4(z) dz \tag{3}$$

$$I_x = I_y = \frac{1}{2}I_z + \pi\rho \int_{z_D}^{z_A} z^2 f^2(z) dz \tag{4}$$

If the curve BC in Figure 1 is irregular, a general solution for an arbitrary SR particle is proposed to calculate these particle properties. A surface mesh can be simply generated in any 3D meshing tool by sweeping the curve BC 360° along the Z-axis. Next step is to divide the minimum axis aligned bounding box (AABB) of the particle mesh into many tiny uniform cells. Normally a cell size of $l/n$, where $l$ is the length of the longest edge in AABB and $n = 100 \sim 200$, will be accurate for the properties calculation. Treating each cell as point mass, we can find $N$ cells whose centers lie within the particle surface mesh. Therefore particle mass $M = Nm$, here $m$ is the cell mass. Center of mass $C = \frac{1}{M}\sum m r_i$, where $r_i$ is the position vector of $i^{th}$ cell center. Principal moments of inertia read $I_x = I_y = \sum m(x_i^2 + z_i^2)$, $I_z = \sum m(x_i^2 + y_i^2)$, here $x_i, y_i, z_i$ are the components of $r_i$.

## 2.2 Treating the contact in 2D space

At the heart of SR-DEM lies the foundational notion that contact detection between a pair of SR particles can be effectively managed within a 2D spatial framework. Employing a master-slave approach, the surface of a master particle undergoes discretization through individual nodes (vertices). Meanwhile, the slave particle is implicitly characterized by means of an explicit and straightforward cross-section or a signed distance field surrounding the cross-section. The validation of Lemma 1 is imperative to establish the legitimacy of the node-to-cross-section contact algorithm.

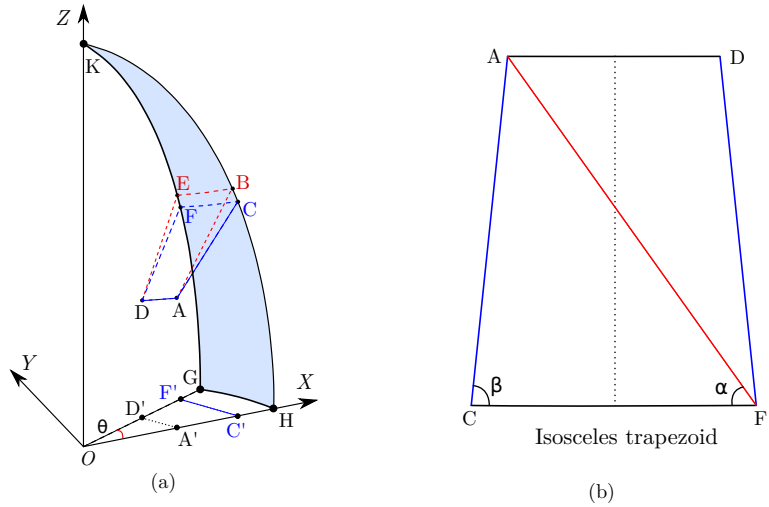

**Figure 3:** Shortest distance from a node to a surface of revolution

**Lemma 1.** *The shortest distance from a node A to a SR, is the shortest distance from A to the cross-section which passes through A and the axis of rotation.*

*Proof.* Assuming the presence of a master node, denoted as A, penetrating the slave surface as depicted in Figure 3, attention is directed towards the planar face *OHK* (comprising the line segments *OH*, *OK*, and the curve *KH*), This planar face constitutes half of the SR particle's cross-section (the curve *KH*) aligned with the axis of rotation (*Z*-axis) and node A. Let C denote the closest node, or one of the closest nodes on the curve *KH* to node A.

The objective is to establish that $d_{AC}$ represents the shortest distance from node A to any other cross-sections (e.g. the curve *KG*) about the axis of rotation. This is substantiated through the following considerations: (1) For an arbitrary node B on the curve *KH*, it is evident that $d_{AC} \leq d_{AB}$; (2) For nodes not lying on the curve *KH*, such as nodes E and F, their positions can be derived by rotating B and C along the curve *KH* with a slight angle $\theta$ around the *Z*-axis. The planar segment *ADFC* is demonstrated to be an isosceles trapezoid, supported by the parallelism of the projected edges $A'D'$ and $C'F'$ in the *XY*-plane. Considering the symmetry about the gray dash-line (axis of symmetry) on the right side of Figure 3, it is evident that the diagonal *AF* is longer than *AC* ($\beta > \alpha$), thereby validating $d_{AE} > d_{AB}$. In summary, $d_{AC} < d_{AF}$ and $d_{AC} \leq d_{AB} < d_{AE}$. Consequently, it is established that the node-to-cross-section distance $d_{AC}$ also represents the node-to-SR distance. (3) Analogously, Lemma 1 remains valid even when the master node A is situated outside the surface of the slave.

$\square$

Given the validity of Lemma 1, it follows that for any master node within the slave surface, the evaluation of the overlap distance from the node to the slave surface can be efficiently executed in a node-to-cross-section manner. This approach exhibits a notable improvement in computational efficiency, several orders of magnitude faster than the conventional node-to-surface algorithm in LS-DEM. The accurate representation of a cross-section necessitates only a few hundred geometric element operations.

## 2.3 Explicit and implicit cross-sections

As the node-to-cross-section contact algorithm is proposed in SR-DEM, a precise description of a SR particle's major cross-section with as few elements as possible is desired an efficient contact detection. In the following the cross-section of a SR particle will be the one through axis of rotation (*Z*-axis) by default, unless a specific one is mentioned.

A SR particle's cross-section can be represented in an explicit or implicit manner. Taking bi-convex tablet [30] as example. It's major cross-section is shown in Figure 4(a): two arcs and two parallel line segments. Since the cross-section is symmetric about the axis of rotation (*Z*-axis), only half of the geometry is needed (face ABCDE); if it is also symmetric about the X-axis, then quarter of the cross-section (face ABCO). This means only half or quarter of the cross-section's geometry information is needed in the node-to-cross-section contact algorithm. Usually a number of 100~200 line segments is adequate for an accurate approximation of the half cross-section, therefore the distance evaluation from a master node to it's slave's cross-section can be efficient by looping over a few hundreds of line segments. Nevertheless, it is not an easy task to check if a master node is inside or outside the slave's cross-section, if only the information of the line segments is given. In addition, a node to a line segment distance calculation involves node projection to the line, and the projected node on the line segment assessment, which is a bit expensive.

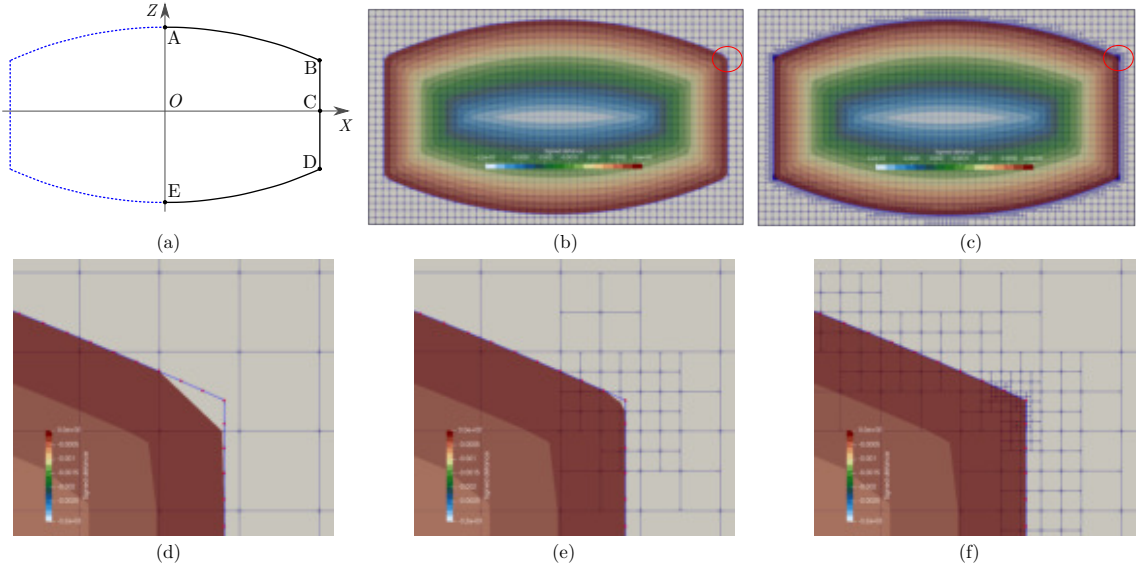

**Figure 4:** The signed distance field of a bi-convex tablet's cross-section. (a) tablet cross-section; (b) signed distance field ($\phi <= 0$); (c) signed distance field ($\phi <= 0$) where the interface cells are refined; (d) closer view of the corner (red circle) in b; (e) grid cells around the corner are refined twice; (f) closer view of the corner (red circle) in c, where the corner sharpness is nearly preserved.

To overcome the drawbacks in the explicit representation, a SR particle's cross-section can be converted into an implicit interface with a signed distance field (SDF) [45]. As shown in Figure 4(b), an uniform grid slightly larger than the cross-section is created. The signed distance of each grid point is obtained by computing the shortest distance from the grid point to the cross-section shown in Figure 4(a), negative inside the cross-section and positive outside. At this point, the implicit cross-section can be described by zero level sets: $S = \{\phi(x) = 0 \mid x \in \mathbb{R}^2\}$, here $\phi(x)$ is the signed distance function. With an intermediate grid resolution where the grid cell size is $l/50$ ($l$: the major axis length of the cross-section), the shape of the cross-section is well reconstructed by the SDF ($\phi <= 0$) in Figure 4(b), except for the sharp corner appears "rounded off" in Figure 4(d). This is because the shape reconstruction introduces artefacts which are a consequence of bilinear interpolation. A sharp corner is identified if the angle is smaller than a threshold (170° here). Sharp corners can be preserved by using multiple channels of SDFs [18], however, complex to implement and computationally expensive in the contact detection.

A simple remedy adopted in this work for "rounded" corners is to refine the grid cells around sharp corners, and build a quadtree for child grid cells within the base (refined) grid cell. A quadtree data structure here helps quickly locate a grid cell (index) where a master node is contained: once the base gird cell is found, the search is narrowed to its child cells. Refining neighboring cells of the sharp corner twice will yield a much smaller round corner of size $l/200$ shown in Figure 4(e), which is an acceptable level of approximation in most cases. For the sake of simplicity, grid cells intersect with or close to the cross-section are termed as interface cells, grid cells contain sharp corners termed as corner cells, and base cells are those who have no parent cell. Even better, thanks to the quadtree built for each refined base cells, a much higher grid resolution around the cross-section can be used with only a small additional cost in locating a cell. In Figure 4(c) the interface and corner cells are refined 2 and 4 times (i.e. level 2 and level 4 refinement respectively), the recovered shape of the cross-section by its SDF is nearly identical to the original one shown in Figure 4(a). A close view of the sharp corner in Figure 4(f) shows that the sharpness of the corner can be considered to be "preserved", because the size of the round corner ($l/n$ where $n = 50 \cdot 2^4 = 800$) in this case is extremely small compared to the length ($l$) of the cross-section's major axis.

## 2.4 Node-to-cross-section contact algorithm

According to Lemma 1, the shortest distance from a master node to the slave's cross-section is equal to the node to surface shortest distance for a pair of SR particles in contact, therefore a node-to-cross-section contact algorithm is naturally the choice for the contact discretization in SR-DEM which is similar to the node-to-surface approach commonly used in FEM for contact simulations [35]. Key difference here is that surface of revolution can be accurately described by the SDF of a SR particle's cross-section.

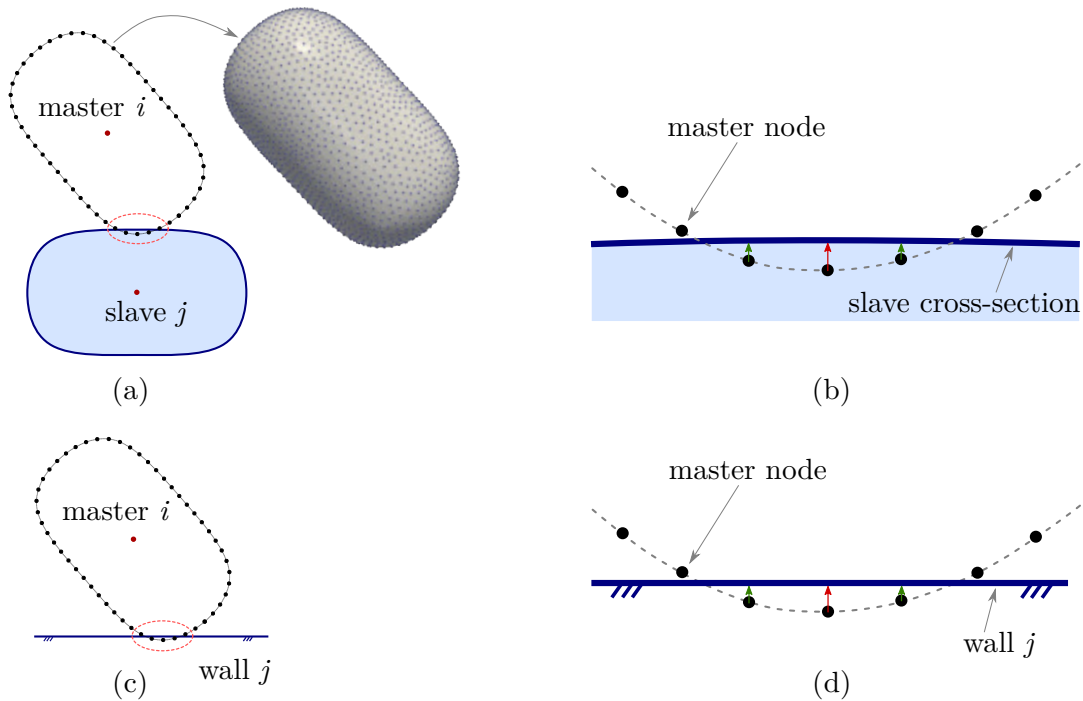

**Figure 5:** Node to cross-section and wall contacts. (a) 2D schematic of two contacting SR particles with master's node-based surface; (b) closer view of the contact in a; (c) 2D schematic of particle-wall contact; (d) closer view of the contact in c.

For a pair of contacting SR particles with global index $i$ and $j$, "master" and "slave" roles need to assigned first. If $i < j$, particle-$i$ will be the master and particle-$j$ the slave in this work. A master particle is described by node-based surface, where these nodes may be simply obtained from the vertices of a 3D triangle mesh shown in Figure 5(a). These nodes distribution on the master can be uniform or adaptive depending on the SR particle's shape: uniform for smooth surface, while adaptive for surface with sharp edges (i.e. more nodes around these edges). The slave's surface is represented by its cross-section that is approximated by line segments in XZ-plane, or the zero level set of the cross-section's SDF as explained previously in section 2.3.

Detailed contact information between master-$i$ and slave-$j$ can be obtained through the broad phase and narrow phase in the node-to-cross-section contact algorithm. The task of the broad phase detection is to check whether a master (surface) node is potentially inside the slave's surface. Let us consider two frames of reference in 3D Euclidean space: *world frame* for global coordinate system and *body frame* for the particle's local (body-fixed) coordinate system. First a master node $A$ in the world frame needs to be transformed to the slave's body frame. Let $\boldsymbol{R}_w$ be the position vector of a master node where the subscript $w$ indicates world frame, $\{\boldsymbol{e}_x, \boldsymbol{e}_y, \boldsymbol{e}_z\}$ be the unit vectors of the slave's principal axes (or local axes) in world frame. Therefore the position vector of a master node in the slave's body frame is $\boldsymbol{R}_b = M\boldsymbol{R}_w$, where the subscript $b$ represents body frame, and the $3 \times 3$ rotation matrix $M$ is simply $[\boldsymbol{e}_x; \boldsymbol{e}_y; \boldsymbol{e}_z]$. Here variables with bold face font indicate vector or tensor. Since the slave's surface is implicitly represented by its half cross-section (X+) in the body frame's XZ-plane (e.g. Figure 4a), position vector $\boldsymbol{R}_b$ is further transformed to $\boldsymbol{R}_{xz} = (x, 0, z)$, here $x = \sqrt{(R_{b,x}^2 + R_{b,y}^2)}$ and $z = R_{b,z}$. If $\boldsymbol{R}_{xz}$ is within the minimum axis aligned bounding rectangle of the slave's cross-section, then the node index of master-$i$ is added into a neighbor list of nodes that are possible inside slave-$j$. Master nodes in the neighbor list will be termed as neighbor nodes in the following.

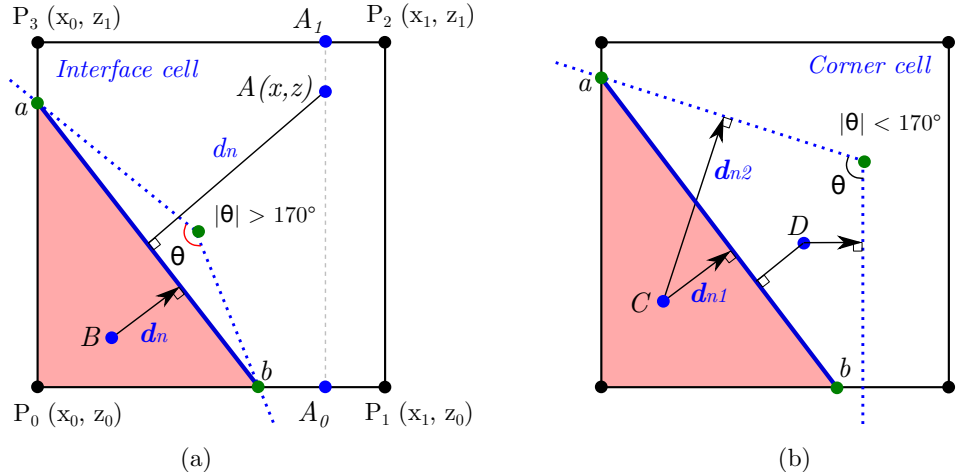

**Figure 6:** Implicit and explicit computation of node to cross-section distance from pre-calculated SDF filed. (a) implicit: bilinear interpolation in interface cell (P0, P1, P2, P3); (b) explicit: distance to line segments in corner cell. Here the blue dashed lines are the line segments of the particle's cross-section. Interface or corner SDF cell is detected with the the threshold angle 170°.

The narrow phase is responsible for the actual distance evaluation from neighbor nodes to the slave's cross-section. Depending on the node density on the master surface, there might be more than one neighbor node "penetrating" into slave's cross-section as illustrated in Figure 5(b).

Bilinear interpolation is chosen in this work for its simplicity and speed, to compute the signed distance ($d_n$) from node to cross-section. Assume a master node $A(x, z)$ is found in an interface cell shown in Figure 6(a), edge $ab$ is the reconstructed line segment of the cross-section from SDF, and triangle $\triangle abP_0$ with red color indicates interior of the cross-section. Here $d_n$ is interpolated from grid points $\{P_0, P_1, P_2, P_3\}$ as follows.

(1) Project node $A$ to cell's bottom and top edges, yields $A_0(x, z_0)$ and $A_1(x, z_1)$;

(2) Get signed distance at $A_0$ and $A_1$ via linear interpolation in the x-direction:

$$\phi(A_0) = \left[(x_1 - x)\phi(P_0) + (x - x_0)\phi(P_1)\right]/(x_1 - x_0) \tag{5}$$

$$\phi(A_1) = \left[(x_1 - x)\phi(P_3) + (x - x_0)\phi(P_2)\right]/(x_1 - x_0) \tag{6}$$

(3) Do the interpolation in the z-direction, yields signed distance $d_n$ at node $A$:

$$\phi(A) = \left[(z_1 - z)\phi(A_0) + (z - z_0)\phi(A_1)\right]/(z_1 - z_0) \tag{7}$$

here $\phi(P_i)$ with $i = 0, 1, 2, 3$ is known signed distance at grid points, and the cell size $L = x_1 - x_0 = z_1 - z_0$. According to relative positions of nodes $A$ and $B$ to edge $ab$ in Figure 6(a), $\phi(A) > 0$ and $\phi(B) < 0$. Here master node $B$ penetrates into the slave's cross-section, therefore its node index and distance vector $d_n$ are added into a list of contacting (master) nodes to the slave. Note that $d_n$ here is computed by bilinear interpolation of pre-calculated distance vectors on grid points, and transformed twice: from XZ-plane (cross-section) to the slave's body frame, and then from the body frame to world frame. Nodes in this list will be termed as contact nodes. In Figure 5(b) there are 3 nodes found inside the slave's cross-section, these contact nodes will be used in the contact force calculation.

As a signed distance field tends to round sharp corners illustrated in Figure 6(b), the actual sharp corner is approximated by an oblique edge $ab$. Therefore there will be some deviation in signed distance interpolation. For example node $D$ will be detected as outside, and the magnitude of distance vector ($d_{n1}$) from $C$ to the cross-section is much shorter than the actual one ($d_{n2}$). This deviation can be significantly reduced by refinement of the corner cells. In this work the corner cell is refined to the size of $l/800$, where $l$ is the length of the cross-section's major axis, which yields very sharp corners as shown in Figure 4(c-f). Alternatively, if the corner cell is not refined, the signed distance can be simply calculated by node to edge distance, as we only need to loop over a few edges around the corner in the slave's body frame XZ-plane, thus still efficient.

Particle-wall contact is handled in the same fashion, where a SR particle is always treated as the master and wall the slave as shown in Figure 5(c-d). A wall can be implicit, e.g. a flat wall defined by $(p, \hat{n})$ where $p$ is a point on the flat wall, and $\hat{n}$ is unit normal vector; or a cylindrical wall defined by $(r, p_1, p_2)$, here $r$ is the radius, and $p_1, p_2$ are two points on the axis of the cylinder. For complex walls, triangular mesh is the common choice, thus a master node to the wall contact is treated as a node to a triangle contact problem. Note that a master node to a wall distance is usually calculated directly. For any master node penetrates into a wall surface, its node index and overlap distance vector is added into the list of contact nodes for the particle-wall pair.

## 2.5 Contact forces

Once the list of contact nodes between master-$i$ and slave-$j$ is built, the normal and tangential components of contact force $F_{i,j}$ can be computed as follows [53].

$$F_{n,ij} = k_{n,ij}\delta_{n,ij} - \gamma_{n,ij}V_{n,ij} \tag{8}$$

$$F_{t,ij} = k_{t,ij}\delta_{t,ij} - \gamma_{t,ij}V_{t,ij} \tag{9}$$

The normal force $F_{n,ij}$ has 2 terms: a contact force ($k_{n,ij}\delta_{n,ij}$) and a damping force ($-\gamma_{n,ij}V_{n,ij}$). The tangential force $F_{t,ij}$ also has 2 components: a shear force ($k_{t,ij}\delta_{t,ij}$) and a damping force ($-\gamma_{t,ij}V_{t,ij}$). Here $k_{n,ij}, k_{t,ij}$ are the normal and tangential contact stiffness; $\gamma_{n,ij}, \gamma_{t,ij}$ are the normal and tangential damping coefficients. $\delta_{n,ij}$ is the pre-calculated distance vector $d_n$ from a contact node to its slave surface, commonly termed as overlap distance vector, and $\delta_{t,ij}$ is an accumulated tangential displacement vector at the contact node. The tangential force will be truncated if it exceeds the magnitude of Coulomb friction $\mu_{ij}|F_{n,ij}|$ where $\mu_{ij}$ is the friction coefficient. For simplicity, the rolling friction is not included in SR-DEM, as it is usually several magnitude smaller than the sliding friction. $V_{n,ij}$ and $V_{t,ij}$ are the normal and tangential components of the relative velocity $V_{ij}$, which can be calculated at the contact node (say $A$).

$$V_{ij} = V_i + \omega_i \times (A - C_i) - V_j - \omega_j \times (A - C_j) \tag{10}$$

where $V_i, V_j, \omega_i, \omega_j$ are the transnational and rotational velocities, $C_i, C_j$ are the center of mass for master-$i$ and slave-$j$.

Coefficients $k_{n,ij}, k_{t,ij}, \gamma_{n,ij}, \gamma_{t,ij}$ depend on the particle's materiel properties as well as the contact condition. In this work non-linear Hertz-Mindlin (HM) model [e.g. 51, 63] is employed.

$$k_{n,ij} = \frac{4}{3}Y^*\sqrt{R^*|\delta_n|} \tag{11}$$

$$\gamma_{n,ij} = \sqrt{5}|\beta|\sqrt{m^*k_{n,ij}} \tag{12}$$

here $Y^*, R^*$ and $m^*$ are the effective Young's modulus, radius and mass. $\beta$ is related to the restitution coefficient $e$, and $\delta_n$ is the signed distance $d_n$ (or overlap) at contact node $A$. In the validation simulations, a simple relationship is used: $k_{t,ij}/k_{n,ij} = 2/7$ and $\gamma_{t,ij}/\gamma_{n,ij} = 1/2$, as the contact dynamics are not very sensitive to precise values of these ratios [53]. For the sake of completeness, the corresponding formulations for coefficients in Eqns. (11-12) are listed below.

$$Y^* = \left(\frac{1-v_i^2}{Y_i} + \frac{1-v_j^2}{Y_j}\right)^{-1}, \quad R^* = \left(\frac{1}{R_i} + \frac{1}{R_j}\right)^{-1}$$
$$G^* = \left(\frac{2-v_i}{G_i} + \frac{2-v_j}{G_j}\right)^{-1}, \quad m^* = \left(\frac{1}{m_i} + \frac{1}{m_j}\right)^{-1}, \quad \beta = \frac{ln(e)}{\sqrt{\pi^2 + ln^2(e)}} \quad (13)$$

where $Y_i, Y_j, G_i, G_j, v_i, v_j, m_i, m_j$ are the Young's modulus, shear modulus, Poisson's ratio and mass of master-$i$ and slave-$j$, respectively. If particles and wall are of the same material, then coefficients $Y^*, G^*$ and $\beta$ are constants.

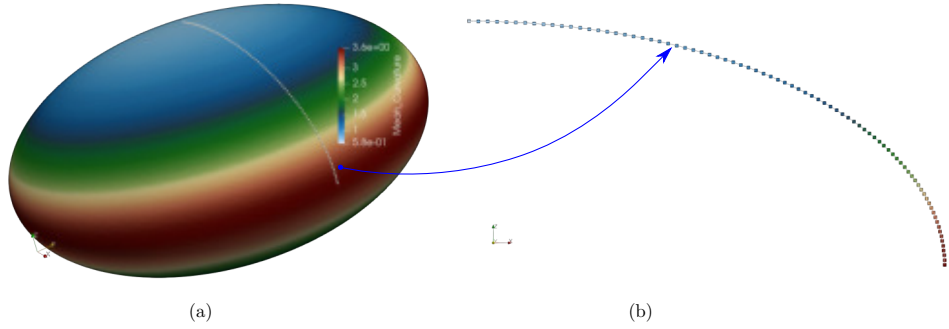

**Figure 7:** SR particle (spheroid) mean curvature. (a) surface mean curvature; (b) mean curvature profile (81 nodes) of the cross-section in 1st quadrant.

To compute effective radius $R^*$, radii of master-$i$ and slave-$j$ ($R_i$, $R_j$) at the contact node $A$, need to be calculated for non-spherical SR particles. Mean curvature $H = \frac{1}{2}(\kappa_1 + \kappa_2)$ at a given point (node $A$) of a surface is used to estimate the radius, here $\kappa_1$ and $\kappa_2$ are the principal curvatures. Thus $R_i$ and $R_j$ can be expressed by $1/H_i$ and $1/H_j$, and $R^*$ reads $(H_i + H_j)^{-1}$. Because of the cylindrical symmetry, any points on a SR particle's surface with same z-coordinate (body frame) has the same curvature, thus we can store pre-calculated surface mean curvature shown in Figure 7(a) on the particle's cross-section. As the particle (spheroid) is symmetric about X- and Z-axis, thus mean curvatures are only needed to be stored on 1/4 of the cross-section (1st quadrant, 81 nodes) shown in Figure 7(b). To get $H_i$ and $H_j$, contact node $A$ is mapped to master and slave body frame, then linear interpolation is used to compute approximated curvature. Note that adaptive nodes on the cross-section is employed here as shown in Figure 7(b), because dense nodes are necessary on steep slopes.

Calculation of torques (i.e. rotational forces) acting on master-$i$ and slave-$j$ is straightforward, as soon as the contact force $F_{ij}$ at contact node $A$ is computed.

$$\text{Torque on master-}i: \quad T_{ij}^{(A)} = (A - C_i) \times F_{ij} \quad (14)$$
$$\text{Torque on slave-}j: \quad T_{ji}^{(A)} = (A - C_j) \times (-F_{ij}) \quad (15)$$

It is important to note that the contact force and torque calculated above are only the contribution from one contact node $A$. If the surfaces of master and slave particles are both convex, then the "deepest point method", often used in super-quadric DEMs, is employed here, such that SR-DEM is less dependent on the node-base surface. Because there will be only one contact point between two convex particles. As shown in Figure 5(b-d), there are 3 contact nodes inside the slave's surface, we only pick the contact node with maximum penetration (the red overlap vector) for the contact force calculation. As the overlap vectors of the penetrated master nodes inside the slave's surface are usually not symmetric about the deepest one, a simple remedy to correct the contact normal may be applied by averaging all of the overlap vectors:

$$\boldsymbol{n} = \sum_{k=1}^{N} \boldsymbol{d}_n^k / |\sum_{k=1}^{N} \boldsymbol{d}_n^k|$$
$$\boldsymbol{d}_n = |d_{n,max}| \cdot \boldsymbol{n}$$
(16)

where $\boldsymbol{d}_n^k$ is the overlap vector (i.e. signed distance vector) of a master node k inside the slave, $d_{n,max}$ is the signed distance of the deepest master node in slave surface, $\boldsymbol{d}_n$ is the resulting overlap distance vector.

On the other hand, if the master or slave or both are non-convex, then we need to account for the contributions of all contact nodes, because there might be more than one contact point between two rigid particles, thus the deepest point method can not be applied here. At this point, the contact force is dependent on the node-base surface. The contact force and torque accounting for all contact nodes are computed as follows.

$$F_{ij} = \sum_{k=1}^{n} F_{ij}^{(A_k)}, \quad T_{ij} = \sum_{k=1}^{n} T_{ij}^{(A_k)}$$
(17)

here $n$ is the number of contact nodes, and superscript $k$ is the local node index of the list of contact nodes. This approach is naturally suitable for rigid convex particles as well, and the contact normal correctness is not needed.

## 2.6 Time integration

Once the contact forces and torques are computed in the narrow phase, particle's transnational and rotational velocities, position and orientation are updated in a small time duration. Equations of motion by applying Newton's second law for a particle-$i$ read:

$$m_i \frac{d\boldsymbol{V}_i}{dt} = \sum_{j=1}^{N_c} (\boldsymbol{F}_{n,ij} + \boldsymbol{F}_{t,ij}) + \boldsymbol{F}_b$$
(18)

$$I_i \frac{d\boldsymbol{\omega}_i}{dt} = \sum_{j=1}^{N_c} (\boldsymbol{T}_{ij} + \boldsymbol{T}_{r,ij})$$
(19)

where $N_c$ is the number of contacting particles to particle-$i$, $\boldsymbol{F}_b$ is particle's body force (e.g. gravity). Rolling friction torque $\boldsymbol{T}_{r,ij}$ is neglected in this work, as rolling friction is usually of several orders of magnitude smaller than sliding friction for particles with relatively smooth surface. As particle's inertia tensor $I_i$ in the world frame is not constant, Eqn. (19) is often computed in the particle's body frame for simplicity.

Second-order Velocity-Verlet algorithm [56] is adopted in this work to integrate the equations of motion (18-19), among other time integration schemes reviewed and compared in literature [11, 32, 52], due to its efficiency, low complexity and acceptable accuracy. The actual implementation of the Velocity-Verlet algorithm in SR-DEM is briefly presented as follows, for the sake of completeness.

$$X(t + \Delta t) = X(t) + V(t)\Delta t + \frac{1}{2}a(t)\Delta t^2 \tag{20}$$

$$V(t + \Delta t) = V(t) + \frac{1}{2}\big[a(t) + a(t + \Delta t)\big]\Delta t \tag{21}$$

where $X(t + \Delta t), V(t + \Delta t)$ are the updated particle position vector and velocity at new time $t + \Delta t$ from known velocity, position and net force at old time $t$. The generalized strategy of the Velocity-Verlet algorithm is implemented in the following sequence.

(1) Calculate velocity by a half time-step: $V(t + \frac{\Delta t}{2}) = V(t) + \frac{1}{2}a(t)\Delta t$;
(2) Update position by a full time-step: $X(t + \Delta t) = X(t) + V(t + \frac{\Delta t}{2})\Delta t$;
(3) Calculate angular momentum by a half time-step: $L(t + \frac{\Delta t}{2}) = L(t) + \frac{1}{2}T(t)\Delta t$;
(4) Calculate angular velocity by a half time-step in body frame: $\omega(t + \frac{\Delta t}{2}) = L(t + \frac{\Delta t}{2})/I$;
(5) Update orientation (in quaternion) by a full time-step: $q(t + \Delta t) = q(t) + [\frac{1}{2}q(t) \circ \omega]\Delta t$;
(6) Update positions of surface nodes based on new particle orientation and center of mass;
(7) Re-compute contact forces and torques from new potential $X(t+\Delta t)$: get $a(t+\Delta t), T(t+\Delta t)$;
(8) Update velocity by 2nd half time-step: $V(t + \Delta t) = V(t + \frac{\Delta t}{2}) + \frac{1}{2}a(t + \Delta t)\Delta t$;
(9) Calculate angular momentum by 2nd half time-step: $L(t + \Delta t) = L(t + \frac{\Delta t}{2}) + \frac{1}{2}T(t + \Delta t)\Delta t$;
(10) Calculate angular velocity by 2nd half time-step in body frame: $\omega(t + \Delta t) = L(t + \Delta t)/I$.

Note that a particle's orientation is described by a 4D vector - quaternion [58] $q(w, v)$ to avoid gimbal lock (e.g. in Euler angles), here $w$ is quaternion's scalar part, and $v(x, y, z)$ is the vector part. An unit quaternion $q$ can be simply constructed from an axis-angle representation $(e, \theta)$: $q(w, x, y, z) = cos(\frac{\theta}{2}) + sin(\frac{\theta}{2})e$, where $e$ is the unit vector of rotational axis, and $\theta$ is the magnitude of the rotation about the axis. In practical implementation, an unit quaternion is also converted to a rotation matrix (Eqn. 22), such that the position transformation between the world frame and the body frame can be done by matrix-vector multiplication. Interestingly, the particle's principal or local axes $\{e_x, e_y, e_z\}$ in the world frame are the columns of the rotation matrix $M(q)$.

$$M(q) = \begin{bmatrix} 1 - 2(y^2 + z^2) & 2(xy - wz) & 2(xz + wy) \\ 2(xy + wz) & 1 - 2(x^2 + z^2) & 2(yz - wx) \\ 2(xz - wy) & 2(yz + wx) & 1 - 2(x^2 + y^2) \end{bmatrix} \tag{22}$$

A proper time-step $\Delta t$ in the time integration is important for a stable and efficient DEM simulation. There are several models to estimate two-particle collision time or so-called critical time step ($\Delta t_c$), which is similar to the Courant number in computational fluid dynamics (CFD) to ensure numerical stability. In this work Rayleigh time $T_R$ based on Rayleigh wave speed [46], and Hertz time $T_H$ based on Hertz theory [20] are employed for the critical time step estimation.

$$\Delta t_c = min\{T_R, T_H\}, \text{ where } T_R = \frac{\pi \overline{R}}{K}\sqrt{\frac{\rho}{G}}, \; T_H = 2.8683\left(\frac{m^{*2}}{R^* \, Y^{*2} \, V}\right)^{0.2} \quad (23)$$

here $Y^*$, $R^*$ and $m^*$ are the same coefficients as defined in Eqn. (13). Coefficient $K$ is a function of Poisson's ratio $\nu$, reads $0.8766 + 0.1631\nu$. $\rho, G$ are the particle's density and shear modulus. For a more conservative estimation, average particle radius $\overline{R}$ is set as minimum particle radius $R_{min}$, and particle's velocity $V$ is the maximum in system. Finally $\Delta t = \Delta t_c / n$, where $n$ is usually set in the range between 10 to 100 to further ensure the numerical stability in DEM simulations.

## 3. Validation cases

To verify the effectiveness of our SR-DEM framework, four validation cases were tested and compared with either analytical or experimental results: (1) particle-wall impact; (2) particle-wall impact; (3) static packing of particles in cylinder; and (4) particles in a rotating drum. The first two cases aim to verify the contact algorithm while the last two for static and dynamic granular systems.

### 3.1 Particle-wall impact

This case aims to validate particle-wall impact as well as the rotational motion using quaternion. As described by Kodam et al. [30], a bi-convex tablet, oriented at a specified angle, impacts a flat wall with a prescribed translational velocity normal to the wall and zero angular speed. The tablet consists of three parts as shown in Figure 8(a) (also in Figure 4a): two cap surfaces (spherical portion) and one cylindrical band. Its size is characterized by the band radius $R_b$, cap height $H_c$ and band height $H_b$. The radius of the cap surface can be derived by: $R_c = \frac{1}{2}(R_b^2 + H_c^2)/H_c$.

If the particle-wall impact is assumed to be frictionless, and the particle's gravity is neglected, the post-impact angular and translational velocities can be written as follows.

$$\omega_y^+ = \frac{mV_z^-(1+e)r_x}{I_y + mr_x^2} \quad (24)$$

$$V_z^+ = \omega_y^+ r_x - eV_z^- \quad (25)$$

where $m$ is the tablet mass, $e$ is the coefficient of restitution at the point of contact, $V_z^-$ is the pre-impact translational velocity (1 $m/s$), $\theta$ is the impact angle, and $I_y$ is the the y-component of the principal moments of inertia. $r_x$ is the projection length of the line segment $OC$ (length $r$) on the X-axis, where $O$ is the tablet center and $C$ is the contact point shown in Figure 8(b). Note that if the impact angle $\theta$ is less than a threshold, say $\theta^* = \sin^{-1}(R_b/R_c) \approx 24.5°$, the impact occurs with the spherical cap, otherwise with the tablet's edge. Therefore $r_x$ reads

$$r_x = \begin{cases} h\cos(\theta) & \text{for } \theta < \theta^* \\ r\sin(\alpha + \theta) & \text{for } \theta \geq \theta^* \end{cases} \qquad (26)$$

here $h$ is the distance from the tablet center $O$ to the center of the sphere $O'$ forming the cap as shown in Figure 8(a). $h = R_c - (H_c + H_b/2)$, $r = (R_b^2 + H_b^2/4)$ and $\alpha = tan^{-1}(\frac{H_b}{2R_b})$. The particle's material properties and simulation parameters are listed in Table 1.

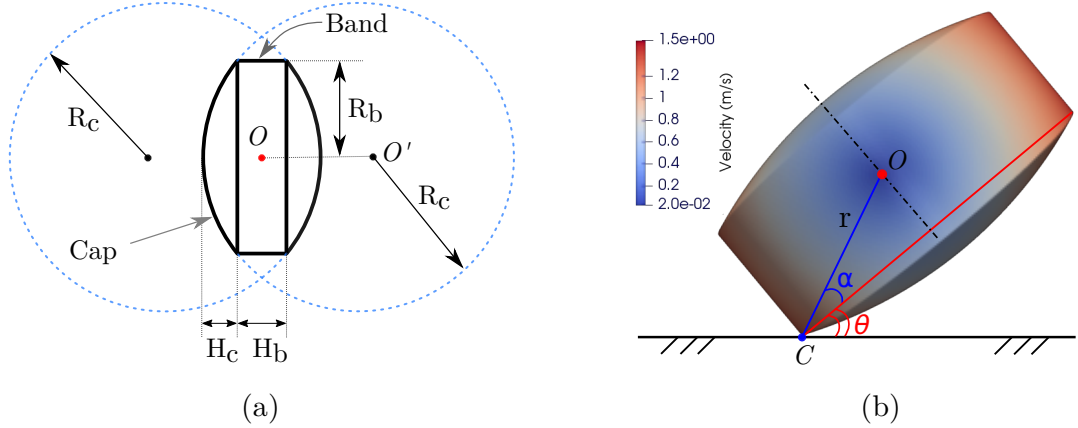

Figure 8: Particle-wall impact. (a) schematic of a bi-convex tablet formed by two spherical portions and a cylinder; (b) schematic of particle-wall impact.

The particle-wall impact is simulated at various orientation angles at an interval of 5°, i.e. $\theta = \{0°, 5°, 10°, \cdots, 90°\}$ with an extra angle at 87.5°. Figure 8(b) shows an instantaneous snapshot of the tablet surface nodes velocity profile just a moment after the particle-wall impact takes place. The velocity of a surface node-$i$ can be calculated by $V_i = V_{cm} + \omega \times r_i$, where $r_i$ is the position vector from the particle center to the contact node-$i$. The post-impact translational and angular

Table 1: Parameters used in particle-wall impact simulation

| Parameter | Value |
| --- | --- |
| Tablet radius $R_b$ (m) | $5.675 \times 10^{-3}$ |
| Tablet band height $H_b$ (m) | $4.0 \times 10^{-3}$ |
| Tablet cap height $H_c$ (m) | $1.23 \times 10^{-3}$ |
| Volume (m³) | $5.312 \times 10^{-7}$ |
| Mass (kg) | $6.328 \times 10^{-4}$ |
| Moments of inertia $I_x$, $I_y$ (kg m²) | $6.231 \times 10^{-9}$ |
| Moments of inertia $I_z$ (kg m²) | $9.396 \times 10^{-9}$ |
| Shear modulus (GPa) | 1.15 |
| Poisson's ratio | 0.3 |
| Coefficient of friction | 0.0 |
| Coefficient of restitution | 0.6 |
| Time step $\Delta t$ (s) | $1.0 \times 10^{-7}$ |

velocities from SR-DEM simulations were compared with the analytical expressions in Figure 10.

For comparison, a secondary tablet shape with slightly rounded edge was also presented. Tablet surfaces are both approximated at relatively fine resolution by 5~6 thousand surface nodes (see Figure 9). Note that the nodes near the rounded edge shown in Figure 9(b) are dense for better shape approximation.

The simulation results from the true bi-convex tablet with sharp edge and the one with rounded edge both have an excellent agreement with the analytical solution. A slight difference in predicted velocities from these two shapes comes from the rounded edge. Note that at impact angles 0°, c.a. 70° and 90°, the tablet's center of mass is right above the contact point, thus the torque generated from contact force is nearly zero. The corresponding (dimensionless) angular and translational velocities are close to 0.0 and 0.6 respectively, which matches the prescribed coefficient of restitution $e$ in Figure 10.

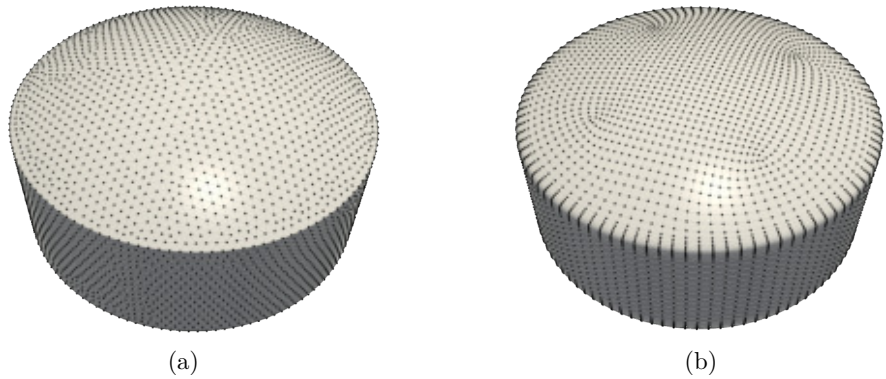

(a)  (b)

**Figure 9:** Tablet shapes used in particle-wall impact simulation. (a) true bi-convex tablet with 5090 surface nodes; (b) bi-convex tablet with rounded edge (6050 surface nodes) where corner radius is $R_b/30$ or $l/60$.

Recall that in section 2.4 the sharp corners of a SR particle's cross-section are handled by two approaches: refinement of the grid cells around the corner, such that the rounded corner is extremely small compared to the particle size; or directly compute the signed distance for nodes near the corner without grid refinement. The simulation results from the rounded tablet (corner radius $R_b/30$) give us a hint that sharp corners might be rounded with a small radius e.g. between $l/50$ and $l/100$, without sacrificing much of the accuracy in shape approximation and contact dynamics. In this way, the "sharp corner" issue in the SDF interpolation may be avoided. In the following simulations, the SDF grid sizes for a SR particle's cross-section are fixed: base cells $l/50$, interface cells $l/200$ and corner cells $l/800$, where $l$ is the length of the major axis of the cross-section. By doing this, a very accurate implicit surface even with sharp edges for slave particles can be ensured.

## 3.2 Particle-particle impact

Normal and tangential contacts between two identical ellipsoidal particles are presented to validate the contact model and the influence of surface curvatures at contact point. Simulation setup is the same to the case of particle-wall impact, the difference here is that the wall is replaced with a fixed particle as shown in Figure 11(a-b). Ellipsoid-2 is initially set right above ellipsoid-1, with a prescribed translational speed ($V_z^-$ = 1 m/s) normal to the X-axis and zero angular speed. Two contact scenarios are tested without considering gravity and friction: head-on contact (Figure 11a) and oblique contact (Figure 11b) where ellipsoid-2's orientation angles are 0° and 45°, respectively. Information about the ellipsoidal particle's geometry, materiel properties as well as the simulation parameters are summarized in Table 2. The velocity profile of Ellipsoid-2 a moment after the impact is plotted in Figure 11(c-d).

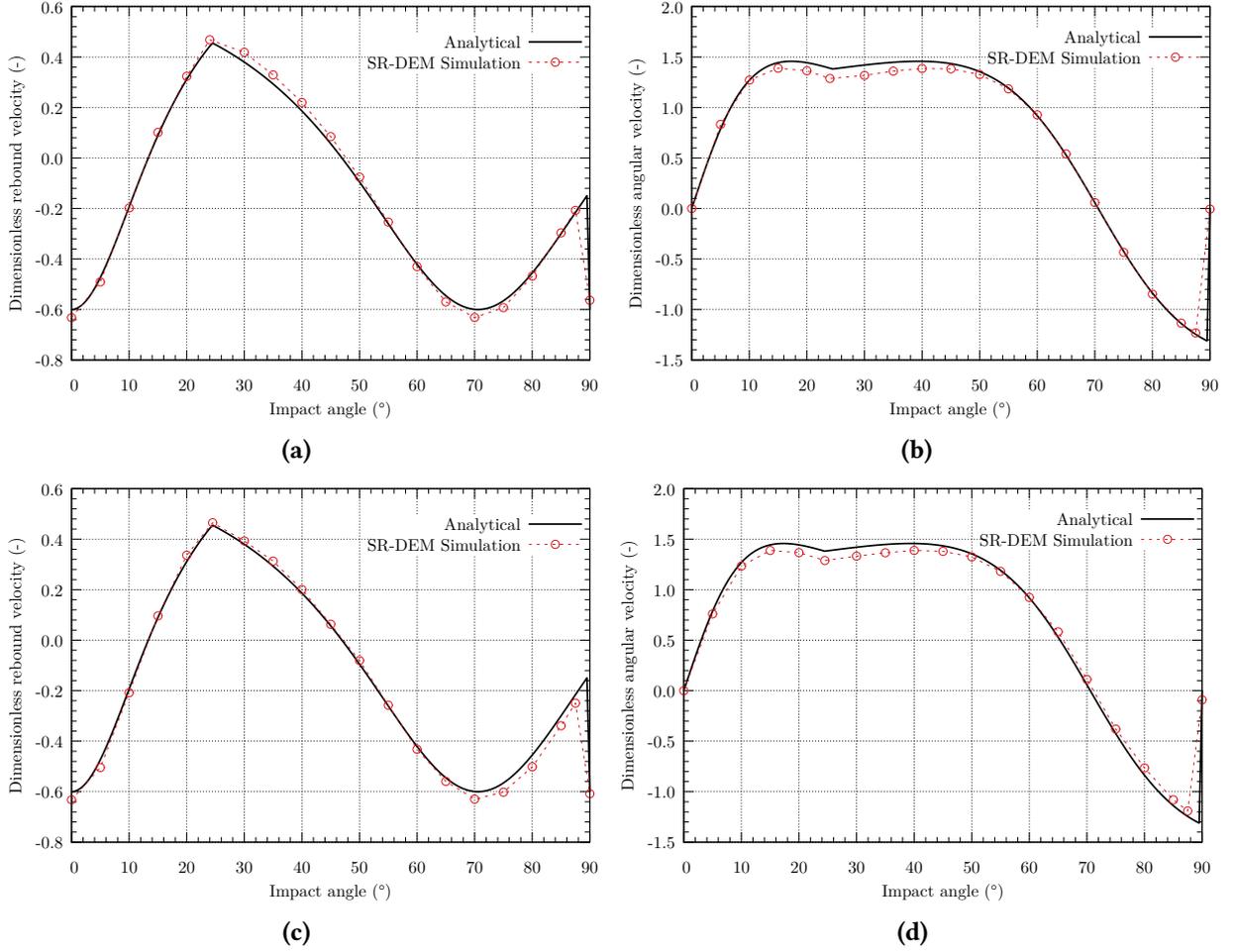

**Figure 10:** Post-impact velocities as a function of impact angle $\theta$ for true and rounded tablets. (a) true tablet: dimensionless translational velocity $V_z^+/V_z^-$; (b) true tablet: dimensionless angular velocity $\omega_y^+ R_b/V_z^-$; (c) rounded tablet: dimensionless translational velocity $V_z^+/V_z^-$; (d) rounded tablet: dimensionless angular velocity $\omega_y^+ R_b/V_z^-$.

Two curvature models were tested in the simulations: the mean curvature from interpolations (see section 2.4) and the onstant curvature from the sphere of equivalent volume. The effective radius ($R^*$) calculated from these curvature models will be termed as mean curvature radius (Rad_mean) and equivalent radius (Rad_eq), respectively. In Figure 12, the normal contact force is plotted as a function of the overlap distance $-\delta n-$ in the range between 0 and 5 μm, and compared with the FEM analysis of head-on contact carried out by Zheng et al. [65].

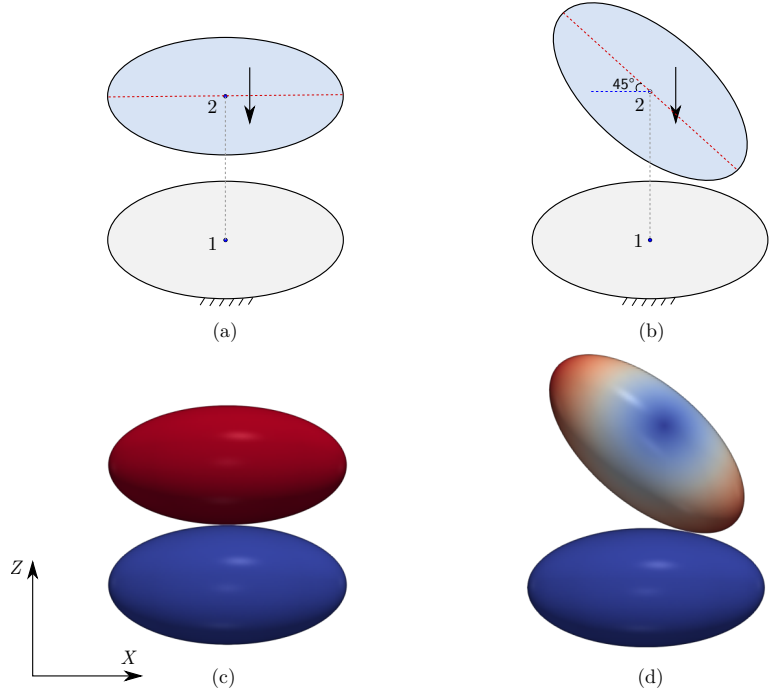

**Figure 11:** Particle-particle impact setup. (a-b) relative position of two ellipsoids before the contact; (c-d) velocity profile of surface nodes a moment after the contact.

**Table 2:** Parameters used in particle-particle impact simulation

| Parameter | Value |
| --- | --- |
| Ellipsoid semi-axes, a, b, c (mm) | 5, 2.5, 2.5 |
| Young's modulus (GPa) | 10.0 |
| Poisson's ratio | 0.3 |
| Density (kg/m³) | 2500.0 |
| Coefficient of restitution (particle-wall) | 0.6 |
| Time step $\Delta t$ (s) | $1.0 \times 10^{-7}$ |

It can be observed that the effective radius from different curvature models has a notable influence on the force-overlap relationship. In the head-on contact case, the predicted normal contact force using the mean curvature radius (blue line) has less deviation to the FEM analysis than the equivalent radius counterpart (bright blue dash-line), while in the oblique contact case the deviation is opposite. Since the normal contact stiffness $k_n$ (see Eqn. 11) is proportional to the square root of the effective radius, and here $R^*_{mean}(0°) > R^*_{eq} > R^*_{mean}(45°)$, it is clear the deviation comes from different contact stiffness in the contact force calculation.

Despite the deviation in the predicted normal contact forces from different curvature models, the rebound velocities of ellipsoid-2 still converge regardlessly, thanks to the non-linear HM contact model. Figure 13 shows the overlap distance and ellipsoid-2's velocity ($V_c$) at the center of mass against time during the entire particle-particle contact. It is obvious that the contact duration varies from different contact conditions: smaller contact stiffness (i.e. smaller $R^*$) leads to longer contact duration as shown in Figure 13(a). Nevertheless, the rebound velocities tend to converge to constant values (see Figure 13b) at the end of the impact regardless of curvature models used in the simulations. It may be concluded that the contact stiffness within certain range does not influence the contact dynamics between particles, even though it does affect the contact duration. To verify this, different Young's modulus values are tested in the range from 10 MPa to 10 GPa, and the conclusion still holds as long as the contact stiffness is not too small. It implies using a small contact stiffness compared to the real value in the contact model will not affect the contact dynamics significantly. This allows us set a relatively large time-step to ensure a fast and stable DEM simulation.

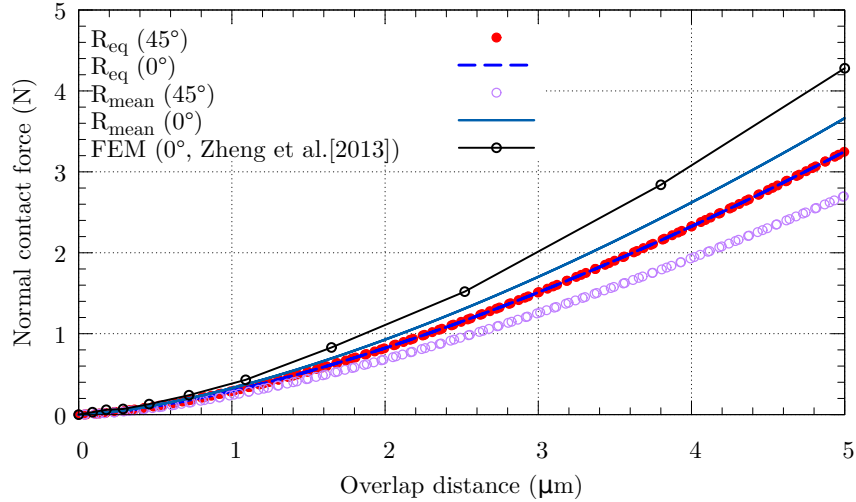

**Figure 12:** Normal contact force vs. overlap distance in head-on and oblique contacts.

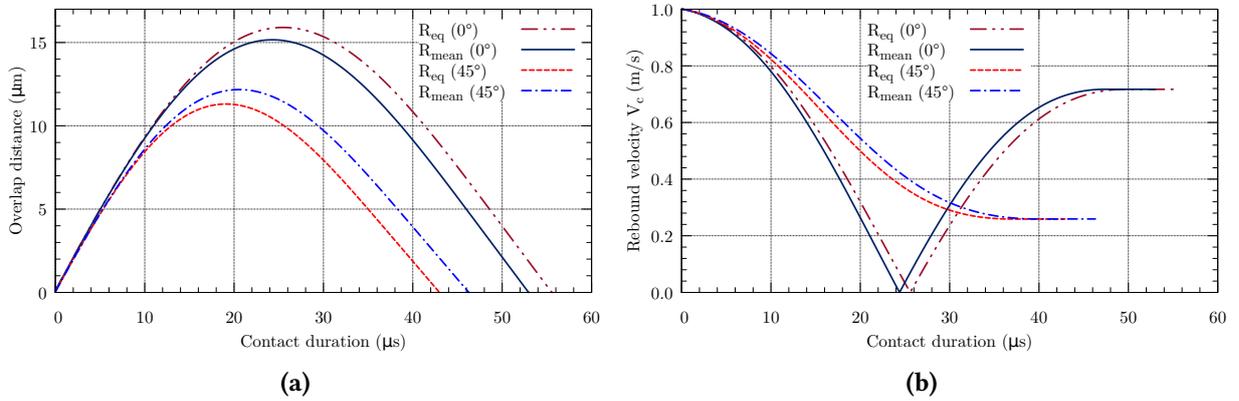

**Figure 13:** Evolution of overlap and particle velocity during head-on and oblique contacts. (a) overlap distance vs. contact duration; (b) velocity of ellipsoid-2 vs. contact duration.

## 3.3 Granular packing in a cylinder

It seems that the proposed contact algorithm in SR-DEM is accurate on the level of a single pair of contacting objects. Next step is to further quantitatively validate SR-DEM against granular systems. In this section static packing of bi-convex tablets and M&M candies in cylindrical containers were simulated and compared with experiments carried out by Kodam et al. [30] and Gao et al. [16] receptively. The tablet shape is identical from the particle-wall impact case (section 3.1), the the M&M candy's shape can be approximately described by an oblate spheroid: $\frac{x^2+y^2}{a^2} + \frac{z^2}{c^2} = 1$ where $a = 6.585$ mm, $c = 3.395$ mm. The information about the cylinder geometry, particle's properties and DEM simulation parameters is summarized in Tablet 3. In the experiments particles are dropped one by one with random orientation and negligible kinetic energy, from random lateral positions at the top of the cylindrical containers. Once the dropped particle comes to rest in containers, the next one is released. In SR-DEM simulations, instead of dropping particles one by one, 3~5 particles with random orientation and position from the same height in experiments, are generated and fall under gravity until they settle in the containers. It was found that the final packing heights in both cases still have good agreement with the experimental results. It implies that we can speed up the simulations by dropping more particles instead of just one at a time.

Table 3: Parameters used in the static packing and rotating drum simulations

| Parameter | Bi-convex tablet | M&M candy |
|---|---|---|
| Particle diameter (mm) | 11.35 | 13.17 |
| Particle thickness (mm) | 6.46 | 6.79 |
| Young's modulus (Pa) | $5.0 \times 10^7$ | $5.0 \times 10^7$ |
| Poisson's ratio | 0.3 | 0.29 |
| Density (kg/m$^3$) | 1191.3 | 1377.0 |
| Volume (m$^3$) | $5.312 \times 10^{-7}$ | $6.162 \times 10^{-7}$ |
| Coefficient of friction | $0.22^{(pw)}, 0.38^{(pp)}$ | $0.3^{(pw)}, 0.3^{(pp)}$ |
| Coefficient of restitution | 0.6 | 0.5 |
| Number of particles | 150 | 250 |
| Cylinder diameter (mm) | 50.6 | 50.8 |
| Cylinder height (mm) | 130.0 | 203.2 |
| Time step $\Delta t$ (s) | $2.0 \times 10^{-7}$ | $5.0 \times 10^{-7}$ |

Superscripts $pw$ and $pp$ denote particle-wall and particle-particle coefficients, respectively.

The final states of the static packing from the experiments and corresponding SR-DEM simulations are compared side by side in Figure 14. The final filling heights in the experiments are $66.87 \pm 2.0$ mm for 150 tablets [30] and $130.0 \pm 1.0$ mm for 250 candies [16], while the measured (average) filling heights in the SR-DEM simulations are $68.5 \pm 2.0$ mm for tablets (yellow color), and $129.0 \pm 1.0$ mm for candies (green color). It demonstrates the SR-DEM's ability to reproduce static granular systems where the characteristic properties like filling height or packing porosity match well against the experimental results. Note that Young's modulus is set as an intermediate value (50 MP) in the simulations instead of using real material properties, because the restitution and Coulomb friction coefficients are the only meaningful parameters that affect the particle dynamics. Therefore relative large time-steps can be used to speed up the SR-DEM simulations.

## 3.4  Dynamic angle of repose

Particles in most particulate processes are often non-static, typical examples include coating, tableting, mixing, chute/belt conveying, crushing, milling, and so on. Therefore it is important to validate the proposed contact algorithm in such dynamic granular systems. In this validation case the flow dynamics of bi-convex tablets in a rotating drum was investigated numerically in SR-DEM simulation, and compared with the experiment carried out by Kodam et al. [30]. The tablet properties and simulation parameters are the same to those used in the static packing of tablets (see Tablet 3). In both experiment and simulation, a drum with an internal diameter of 100 mm and a length of 56 mm contains 150 tablets, and is tumbled at a rotational speed of 25 rpm.

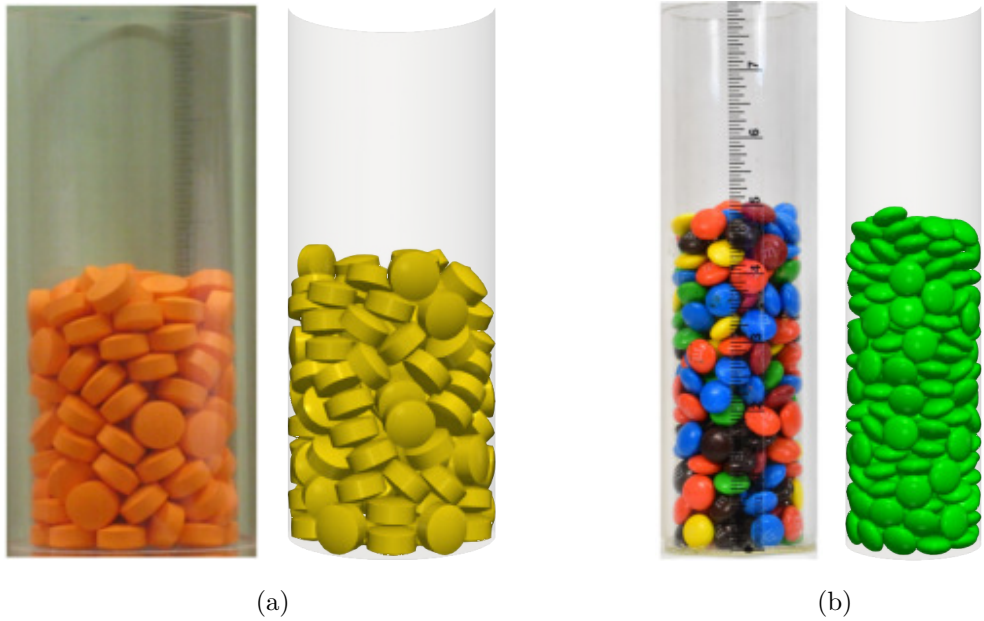

(a)  (b)

**Figure 14:** Static packing of (a) 150 bi-convex tablets, and (b) 250 M&M candies in cylindrical containers. Here left are experiments [16, 30] and right are the SR-DEM simulations in (a-b). Note that the cylinder (130 mm) in (a) are scaled such that its height is equal to the cylinder (203.2 mm) in (b).

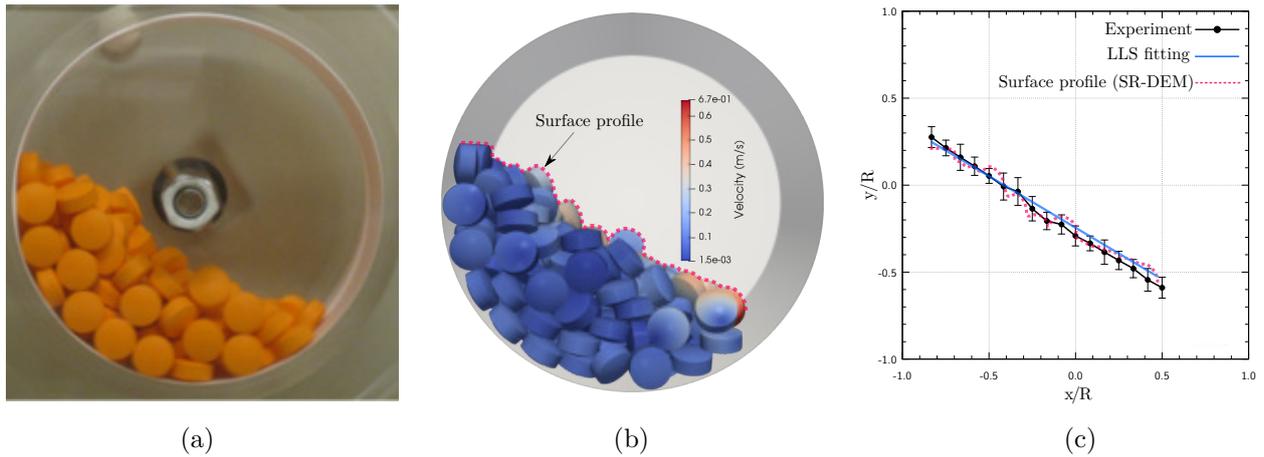

(a)  (b)  (c)

**Figure 15:** Snapshots of bi-convex tablets in a rotating drum from (a) experiment by Kodam et al. [30], and (b) SR-DEM simulation with a dotted curve surface profile. (c) averaged free surface in experiments and linear least squares (LLS) data fitting of the surface profile in (b). Note that the free surface point positions here are dimensionless where $R$ is the drum radius.

Tbe snapshots of the tablets bed in the rotating drum from the experiment and SR-DEM

simulation are compared in Figure 15(a-b), an overall good agreement between the experiment and the SR-DEM simulation can be observed. Due to the fluctuation of the particle bed, the free surface profile is usually not linear as shown in Figure 15(b). To estimate the dynamic angle of repose, ten snapshots (view direction - drum axis) were taken at one second intervals same to the experiment. For each snapshot a free surface profile (2D curve) can be determined, and a linear least squares (LLS) data fitting is used to evaluate the angle of repose. A number of 50~100 control points evenly distributed on the profile curve may be adequate for the LLS fitting. A subsequent averaging over all individual snapshots will render the dynamic angle of repose in the rotating drum. In the experiment [30] the pixel positions of the free surface at regular horizontal intervals were averaged over 10 images as shown in Figure 15(c), and then fitted with a straight line to determine the surface dynamic angle of repose.

The mean dynamic angle of repose from SR-DEM simulations is 31.7 ± 2.0°, while 33.1 ± 3.3° in the experiment. SR-DEM slightly under-predicts the angle of repose, but still lies within the uncertainty interval of the experimental result. This validation case further demonstrates the effectiveness of SR-DEM in modeling dynamic granular system.

## 4. Optimal resolution of node-based surface

To model granular systems with non-spherical particles in DEMs, realistic particle shapes need to be approximated properly for a balance between the computational efficiency and accuracy. For the node-to-cross-section contact algorithm, we want the number of surface nodes as few as possible, as long as the bulk granular properties such as packing porosity, static or dynamic angle of repose, and flow patterns, etc., from DEM simulations are close to the experimental results. Therefore, an optimal particle surface resolution needs to be estimated prior to the actual DEM simulations, which usually takes few days to few weeks for a few dozens to hundreds of thousand particles. One possible approach to find the optimal resolution is to perform a "mini-version" of the experiment numerically. The particle resolution is then gradually increased until the major bulk granular properties converges. At this point, we may say the optimal particle surface resolution is found.

For simplicity, a rotating drum with particles inside will be used here to showcase the process to find optimal resolution. The simulation setup is identical to the validation case in section 3.4, except for the 150 bi-convex tablets being replaced by 130 M&M candies (see Tablet 3) such that the total volume of particles is close. The surface resolutions are set as $500 \times 2^n$ nodes (n = 0, 1, 2, 3, 4) with an extra intermediate resolution at 3000 surface nodes. Figure 16 shows the node-based surface models and the corresponding snapshots of particle beds in the rotating drum. For each case 10 snapshots are taken every half second after the particle beds are more or less stable dynamically, then post-processed the same way in section 3.4 to estimate the averaged dynamic angles of repose (DAoR).

As shown in Figure 17, the averaged DAoR (30.5°) at the coarsest resolution of 500 nodes is noticeably larger than other cases, it is because the surface is not smooth but a bit "blocky" (see Figure 16a), thus the interlocking between particles is increased. When we double the number of surface nodes to previous resolution (except for 3k nodes), the DAoR is quickly approaching to a converged value around 29.5°. The slight fluctuation on the DAoR values may come from the uncertainty of snapshot sampling. The results suggest that a minimum resolution of 1000~2000 surface nodes may be adequate for a relatively accurate DEM simulation with least computational effort.

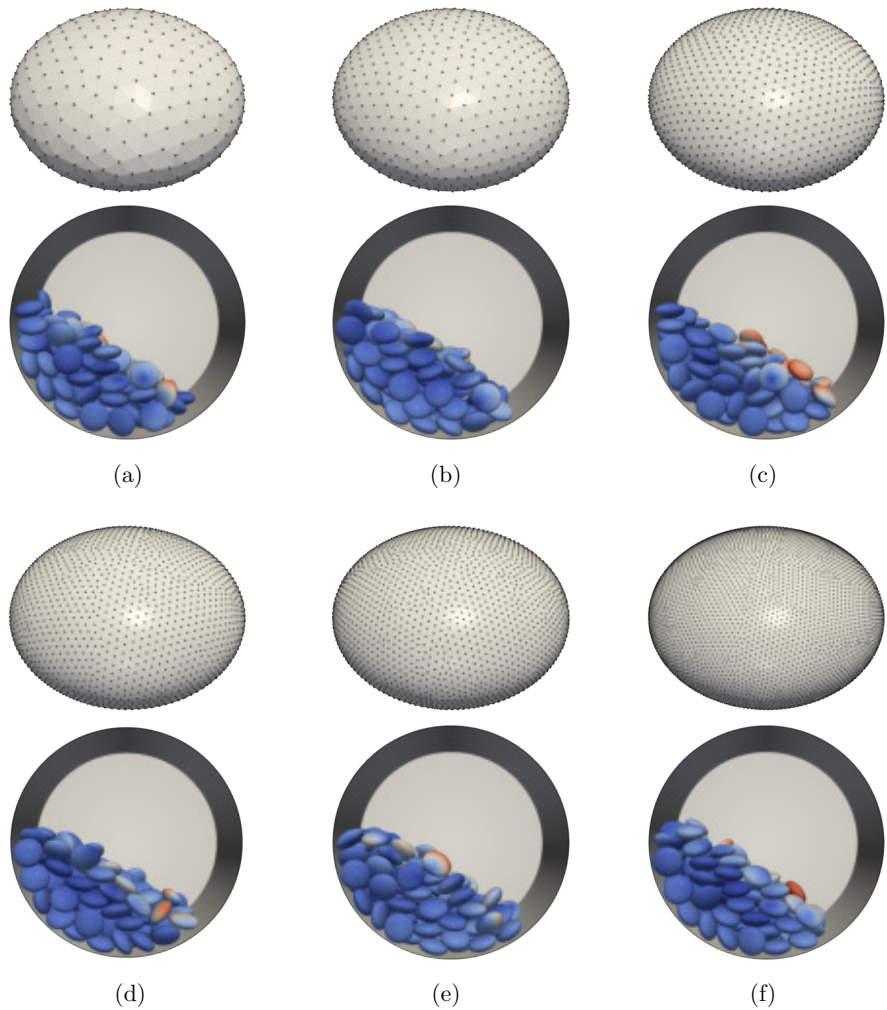

**Figure 16:** SR-DEM simulation of 130 M&M candies in a rotating drum at a rotational speed of 25 rpm, with increasing particle surface resolution in number of nodes: (a) 500; (b) 1000; (c) 2000; (d) 3000; (e) 4000 and (f) 8000. In each sub-figure the top is particle's node-based surface and bottom is the corresponding snapshot of the particle bed (colored by surface node velocity magnitude).

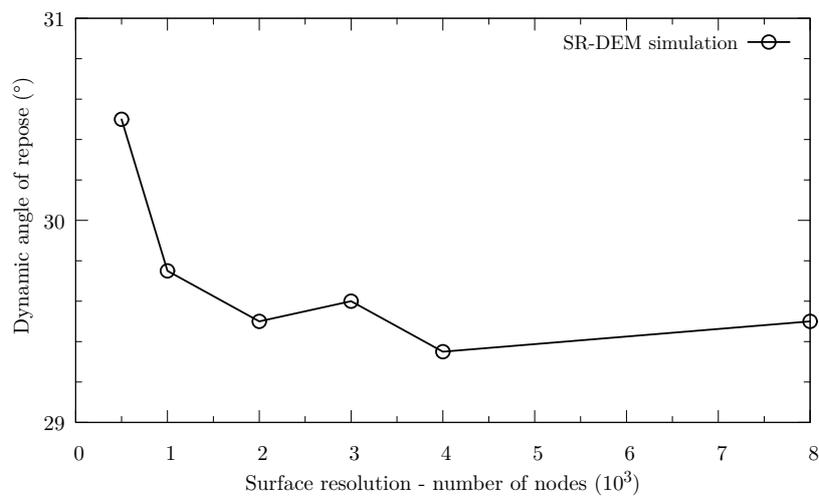

**Figure 17:** Averaged dynamic angles of repose with increasing surface resolution

For the case with 130 bi-convex tablets in section 3.4, the averaged DAoR is 31.7° from the SR-DEM simulation. It implies that the particle shape is the major factor that affects the averaged DAoR, as the sizes of bi-convex tablet and M&M candy are close. It is also important to note that the optimal resolution is only for one particle shape and size. For particles with different shape and size we may need to repeat the process.

## 5. Discussion and concluding remarks

In this work, a novel SR-DEM framework is developed to simulate systems of particles with surface of revolution. Taking advantage of the unique geometry feature of SR particles, the complex particle-particle and particle-wall contacts can be handled with an efficient node-to-cross-section contact algorithm. The proposed framework is validated by various aspects: a particle impact on a wall and another particle to verify the contact algorithm in detail; granular packing in cylindrical containers, and particles in rotating drums with test the SR-DEM's ability to reproduce static and dynamic granular systems. The results from the SR-DEM simulations showed an overall satisfactory agreement with either analytical and experimental data from literature, which demonstrate SR-DEM's capability to capture the behavior of granular systems. It is also suggested to find an optimal particle surface resolution prior to serious DEM simulations, in order to achieve an acceptable accuracy with least computational effort. For SR particles with smooth surface 1000~2000 nodes may be adequate to capture the dynamic behaviors of particles in a rotating drum. In the case of particles with sharp edges a relatively finer resolution might be necessary.

For SR particles with simple cross-section that consists of a few line segments, arcs or any mathematical curves, the signed distance field (SDF) of the particle's major cross-section is not necessary as the node distance to these curves can be directly computed. For example, the cross-section of bi-convex tablet is composed of two arcs and two line segments, M&M candy's cross-section is simply an ellipse. Thus we only need to loop over a master node to these implicit curves few times, instead of looping over few hundreds of SDF cells. This implies SR-DEM has the potential to simulate SR particles with simple cross-section very efficiently.

References 31[50] Rakotonirina, A. D., Radjai, J.-Y. D. F. and Wachs, A. 'Grains3D, a flexible DEM approach for particles of arbitrary convex shape – Part III: extension to non-convex particles modelled as gluedconvex particles'. In: *Computational Particle Mechanics* 6 (2019), pp. 55–84.

[51] Renzo, A. D. and Maio, F. P. D. 'Comparison of contact-force models for the simulation of collisions in DEM-based granular flow codes'. In: *Chemical Engineering Science* 59 (2004), pp. 525–541.

[52] Rougier, E., Munjiza, A. and John, N. W. M. 'Numerical comparison of some explicit time integration schemes used in DEM, FEM/DEM and molecular dynamics'. In: *International Journal for Numerical Methods in Engineering* 61 (2004), pp. 856–879.

[53] Silbert, L. E. et al. 'Granular flow down an inclined plane: Bagnold scaling and rheology'. In: *Physcial Review E* 64 (2001).

[54] Sinnott, M. D. and Cleary, P. W. 'The effect of particle shape on mixing in a high shear mixer'. In: *Computational Particle Mechanics* 3 (2016), pp. 477–504.

[55] Song, Y., Turton, R. and Kayihan, F. 'Contact detection algorithms for DEM simulations of tablet-shaped particles'. In: *Powder Technology* 161 (2006), pp. 32–40.

[56] Swope, W. C. and Andersen, H. C. 'A computer simulation method for the calculation of equilibrium constants for the formation of physical clusters of molecules: Application to small water clusterss'. In: *The Journal of Chemical Physics* 76 (1982).

[57] Wang, S. and Ji, S. 'Poly-superquadric model for DEM simulations of asymmetrically shaped particles'. In: *Computational Particle Mechanics* 9 (2022), pp. 299–313.

[58] Wikipedia contributors. *Quaternions and spatial rotation.* [Online; accessed 06-May-2022]. 2022. URL: https://en.wikipedia.org/wiki/Quaternions_and_spatial_rotation.

[59] Yu, F., Zhang, S., Zhou, G., Zhang, Y. and Ge, W. 'Geometrically exact discrete-element-method (DEM) simulation on the flow and mixing of sphero-cylinders in horizontal drums'. In: *Powder Technology* 336 (2018), pp. 415–425.

[60] Yuan, F.-L. 'Combined 3D thinning and greedy algorithm to approximate realistic particles with corrected mechanical properties'. In: *Granular Matter* 21 (2019).

[61] Zhan, L., Peng, C., Zhang, B. and Wu, W. 'A surface mesh represented discrete element method (SMR-DEM) for particles of arbitrary shape'. In: *Powder Technology* 377 (2021), pp. 760–779.

[62] Zhao, H.-K., Osher, S. and Fedkiw, R. 'Fast surface reconstruction using the level set method'. In: *Proceedings IEEE Workshop on Variational and Level Set Methods in Computer Vision.* 2001, pp. 194–201.